\documentclass[11pt]{article}

\usepackage[a4paper,lmargin=2.5cm,rmargin=2cm,tmargin=1cm,bmargin=1cm,includehead,includefoot]{geometry}

\usepackage{graphicx,amsmath,url}
\usepackage{microtype,lmodern}
\usepackage{mathtools,amssymb,gensymb}
\usepackage{tgtermes}
\usepackage[T1]{fontenc}
\usepackage[utf8]{inputenc}

\usepackage{acronym}
\usepackage{amsmath}
\usepackage{array}
\usepackage{bbold}
\usepackage{booktabs}
\usepackage{dsfont}
\usepackage{fancyhdr}
\usepackage{relsize}
\usepackage[font={small,up,singlespacing}]{subcaption}
\usepackage[backend=bibtex,style=chem-angew,sorting=none,autocite=superscript]{biblatex}
\usepackage{pdfpages}
\usepackage{bbm}
\usepackage{color}
\usepackage{siunitx}
\usepackage{braket}
\usepackage{mwe}
\usepackage{biblatex}
\usepackage{multirow}
\usepackage{mathtools}
\usepackage{xcolor}
\usepackage{float}
\usepackage{colortbl}
\usepackage{siunitx}\DeclareSIUnit\molar{\textsc{M}}

\usepackage[colorlinks,breaklinks]{hyperref} 
\usepackage{cleveref}

\hypersetup{hypertexnames=true, linkcolor=blue, anchorcolor=black, citecolor=blue, urlcolor=blue}
\graphicspath{{./}{imglocal/}{img/}}
\title{\textbf{Hybrid Classical/Machine-Learning Force Fields for the Accurate Description of Molecular Condensed-Phase Systems}}

\author{Moritz Th\"urlemann$^a$, Sereina Riniker$^a$*}
\date{$^a$~Department of Chemistry and Applied Biosciences, ETH Zurich, Vladimir-Prelog-Weg 2, 8093 Zurich, Switzerland. *Email: sriniker@ethz.ch.}

\begin{document}

\maketitle

\section*{Abstract}
Electronic structure methods offer in principle accurate predictions of molecular properties, however, their applicability is limited by computational costs. Empirical methods are cheaper, but come with inherent approximations and are dependent on the quality and quantity of training data. The rise of machine learning (ML) force fields (FFs) exacerbates limitations related to training data even further, especially for condensed-phase systems for which the generation of large and high-quality training datasets is difficult.
Here, we propose a hybrid ML/classical FF model that is parametrized exclusively on high-quality \textit{ab initio} data of dimers and monomers in vacuum but is transferable to condensed-phase systems. The proposed hybrid model combines our previous ML-parametrized classical model with ML corrections for situations where classical approximations break down, thus combining the robustness and efficiency of classical FFs with the flexibility of ML. 
Extensive validation on benchmarking datasets and experimental condensed-phase data, including organic liquids and small-molecule crystal structures, showcases how the proposed approach may promote FF development and unlock the full potential of classical FFs.


\section{Introduction}
An accurate description of the physical interactions between atoms in condensed-phase molecular systems remains one of the biggest challenge in computational chemistry. 
Electronic structure methods are in principle able to describe properties of such systems reliably~\cite{MP2Water, SCANWater}. However, access to long time-scales and large systems is severely limited by the associated computational cost. Due to the computational complexity of electronic structure methods, this issue is unlikely to be resolved solely by additional computational power in the near future~\cite{QuantumComplexity}.  
As a solution, approximate methods, such as force fields (FFs)~\cite{FixedChargeFF} or semi-empirical quantum chemistry methods, have been developed~\cite{PopleSemiEmpirical, XTB}. Especially FFs enable routine access to large systems at microsecond time scales~\cite{Anton3}. However, approximations inherent to FFs and semi-empirical methods limit their ability to describe certain interactions, for instance polarization~\cite{ACHC}.

With the development of machine learning (ML) potentials during the last decade (see e.g. Refs.~\cite{HDNNP, SCHNET, SOAP, MLPotentials}), a new paradigm has emerged for the computational study of atomic systems. Thanks to the fast-paced development of underlying architectures, ML potentials achieve now routinely errors on training sets and validation sets comparable to the errors of the reference method itself~\cite{GDML, GemNet, SpookyNet, PAINN, Allegro, MACE, E3DesignSpace}. However, existing models are still limited by their robustness for long prospective simulations, transferability, and computational cost~\cite{StabilitySimulations, MLAdvancesandChallenges}. Especially the ability to transfer from small systems in vacuum, i.e., monomers and dimers, to diverse condensed-phase systems has, to our knowledge, not been demonstrated yet. In practice, extending the sampling of accurate electronic structure methods with ML could be one of the most interesting use cases for ML potentials~\cite{BehlerWater, HutterWater}.

Transferability from the gas phase to the condensed phase is essential due to the computational cost associated with the generation of large training sets with highly accurate reference methods. With increasingly accurate ML models, the quality of the reference method becomes decisive as the model itself will no longer be the leading error source. As an exemplary use case, special attention is given to molecular crystals in this work. Crystal structure prediction (CSP), i.e., the prediction of the spatial arrangement of atoms in the crystalline phase given a chemical or structural formula, is a long-standing challenge in physical sciences~\cite{CSP, CSP1, CSPNature, CSP2}. 
As demonstrated in the sixth CSP blind test~\cite{CSP6}, successful prediction and ranking of crystal structures does not only hinge on the ability to accurately predict the lattice energy. Instead, the importance of entropic contributions, and possibly to a lesser degree nuclear quantum effects, has emerged~\cite{LatticeVibrations, HojaCSP, AnharmonicFE0, AnharmonicFE1, AnharmonicFE2}. Obtaining a good estimate of these contributions requires, however, extensive sampling.

In this study, we build on the developments and results proposed in previous work and extends the formalism proposed in Ref.~\cite{MLFF}. As the most important addition, we introduce a ML-parametrized two-body potential to improve the description of short-range interactions. This two-body potential incorporates directional information through the use of static multipoles and induced dipoles. Such generic ML n-body potentials could greatly facilitate the development of potentials for situations where classical approximations break down or in cases where the derivation of an analytic functional form is difficult. At the same time, interpretability is retained to a large degree. 

In this work, particular emphasis is put on the transferability from small and isolated systems to large systems in the condensed phase. We argue that this size-transferability provides not only a strong signal that the model predicts interactions in accordance with underlying physical laws, but also enables parametrization on high-quality data which is typically only available for small systems. At present, size-transferability is possibly the most overlooked property for ML potentials, which are either only trained and applied to small systems where such effects are not apparent, or which are only trained and applied to condensed-phase systems, possibly obscuring this limitation.
To achieve this goal, the proposed model relies on existing classical models, which describe the interactions between atoms where possible, such as classical dispersion models or multipole electrostatics. ML comes into play to (i) parametrize these classical models, and (ii) to replace and correct the classical description.
The former takes advantage of the automatic differentiation based parametrization framework described in previous work~\cite{MLFF}. 
Automatic differentiation has emerged as a powerful tool in computational science, permitting efficient gradient-based parametrization of physical models~\cite{TBAD, QDC, JAXMD}.
The latter is used to introduce a higher degree of flexibility, which is necessary for situations where classical approximations break down, for instance at short distances and large overlaps.


\section{Theory}
\subsection{Model Overview}
We assume a classical description of atomic interactions. Within this formalism, molecules are described as graphs with nodes corresponding to atoms and edges to covalent bonds. This notion allows for the definition of learned atom types following the formalism that we proposed in our previous work on graph neural network (GNN) parametrized FFs~\cite{MLFF}. At the same time, the classical description permits a separation into \textit{inter}molecular and \textit{intra}molecular interactions. 

\begin{figure}[H]
    \includegraphics[width=\textwidth]{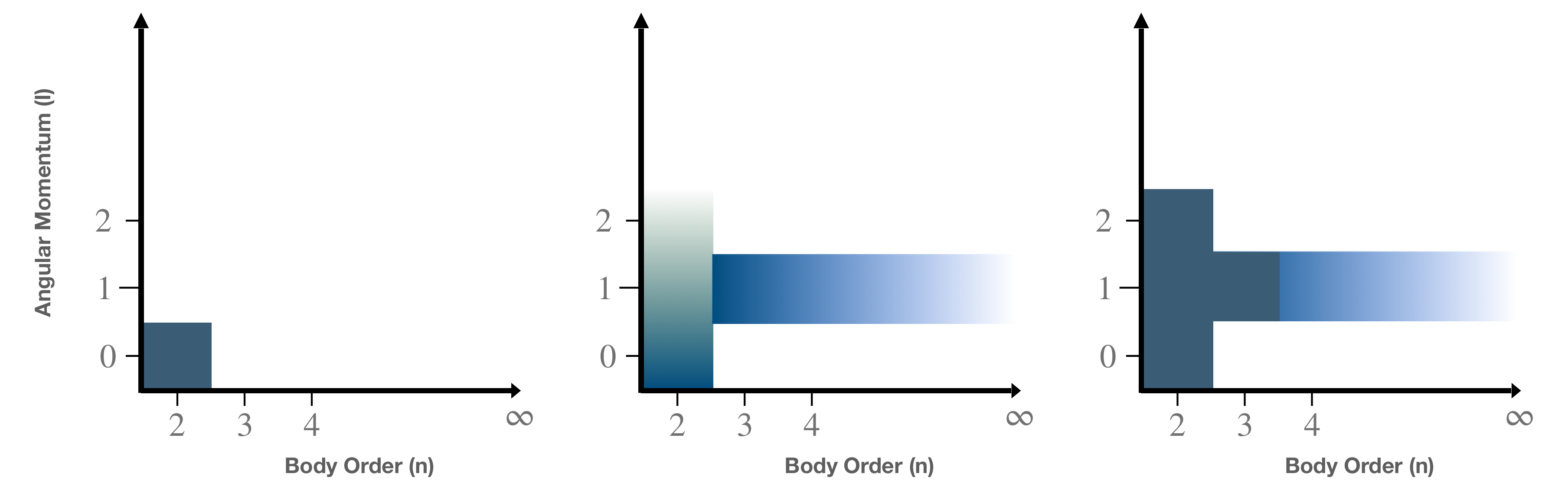}
    \caption{Classification of classical intermolecular interactions. (Left): Classical fixed-charge FF with point charges and a Lennard-Jones potential. (Middle): Classical polarizable FF such as AMOEBA~\cite{AMOEBA} with the additional inclusion of polarization ($\infty$, 1) and atomic multipoles (2, (0, 1, 2)). (Right): Model proposed in this work, which includes a three-body dispersion term (3, 1) and a pairwise ML potential (2, (0, 1, 2)) compared to the polarizable FF. The pairwise ML potential can account for directional interactions in a systematic manner.}
\label{fig:FFCategorization}
\end{figure}

Taking advantage of this separation, an intramolecular potential was parametrized on energies, gradients, and multipoles (MBIS~\cite{MBIS}) of isolated molecules on {PBE0}/def2-TZVP level of theory~\cite{PBE0, PBE, DEF2}. For the treatment of intermolecular interactions, an additional separation of long-range and short-range interactions is introduced. We assume that long-range interactions, including electrostatics, polarization, and dispersion, are accurately captured by classical models using atomic multipoles and polarizabilities~\cite{Applequist, Thole}, and the D3 dispersion correction~\cite{D3, D3BJ}. As these descriptions break down at short distances, a number of classical models have been put forward in recent years ~\cite{HIPPO, AnisotropicExchange}, which resolve this limitation, for instance through the use of charge-penetration models for the description of short-range electrostatics~\cite{AMOEBAChargeDamping}.
Here, a pairwise ML potential is adopted as an alternative.
Within a classical formalism, potentials can be classified according to the information used as input feature (Figure \ref{fig:FFCategorization}). We follow a classification based on two fundamental dimensions: The degree of directional information (angular momentum) and the number of particles (many-body order) involved in the interaction.

Thanks to their flexibility, ML potentials can be parametrized in a systematic manner according to the proposed categorization.  
In this work, we limit ourselves to an anisotropic pairwise ML potential, which is applied to intermolecular atom pairs at short distances in addition to dispersion, electrostatic, and polarization interactions. We will refer to this model as ANA2B, i.e., an anisotropic, non-additive FF in combination with a two-body ML potential.

As input features, pairwise distances, atom types, and the interaction coefficients of static and induced multipoles are used. A more detailed description of these features is given in Section~\ref{sec:short_range}. The pairwise intermolecular interaction is trained on neutral systems of the DES5M dataset~\cite{DEShawDimers}, which includes intermolecular potentials of small molecule dimers obtained with spin-network-scaled MP2 (SNS-MP2)~\cite{SNSMP2, SNS, MP2}. At present, DES5M is the largest dataset of high-quality intermolecular interactions. Since datasets of similar quality are not available for condensed-phase systems, we limit ourselves to DES5M for intermolecular interactions of dimers and {PBE0}/def2-TZVP for intramolecular interactions of monomers (see Section~\ref{sec:intra}), with the aim to develop a model that can transfer from these small systems to the condensed phase.

\subsection{Molecular Graphs and Atom Types}\label{sec:molecular_graphs}
The notion of atom types used as part of the proposed model relies on the formalism proposed in Ref.~\cite{MLFF}. This formalism makes use of atom types, which are learned from molecular graphs, i.e., graphs that do not include information about the geometry of a molecule but only its covalent bonds. Graphs were constructed in the same manner as described in Ref.~\cite{MLFF}. We will refer to these molecular graphs as $\mathcal{G_{\text{Mol}}}$.
Atom types extracted from these molecular graphs are used as input features for subsequent tasks. The atom type of atom $i$ is defined as an $n$-dimensional feature vector $h_i^n\in \mathbf{R^n}$, where the superscript $n$ indicates the order, i.e., $h^0$ corresponds to the element itself, $h^1$ to an atom type that incorporates information about the immediate neighbours, and so on. Atom types are learned as part of the training process with a message passing GNN as proposed in Ref.~\cite{InteractionNetwork}. 

\subsection{Geometric Graphs}\label{sec:geometric_graphs}
The models for the prediction of atomic multipoles and the correction to the intramolecular potential $V_{\Delta \text{ML}}$ use geometric information. These graphs were constructed by including an edge between all atoms, which were $<5\,$\AA~apart. Following the approach described by Gasteiger \textit{et al.}~\cite{DIME}, distances were encoded with $20$ Bessel functions and enveloped with a cutoff function to ensure a smooth cutoff. Element types were encoded as one-hot vectors serving as initial node features.

\subsection{Message Passing Graph Neural Networks}\label{sec:message_passing}
Given a molecular or geometric graph $\mathcal{G}=(V, E)$ with nodes $V$ and edges $E$ as described above, message passing can be defined as~\cite{GilmerQuantumGNN, GNN},
\begin{equation}
    \begin{aligned}
        h_i^{l+1} &= \phi_h(h_i^l, \sum_{j\in N(i)}\phi_e (h_i^l, h_j^l, u_{ij})) ,
    \end{aligned}
    \label{eq:message_passing_c6}
\end{equation}
where $h_i^l\in \mathbf{R^n}$ describes the hidden-feature vector of node $v_i$ after $l$ iterations, $u_{ij}\in \mathbf{R^n}$ the edge feature of edge $e_{ij}$ between node $i$ and $j$, and $N(i)$ denoting the set of neighbours of $v_i$. $\phi_e$ and $\phi_h$ refer to edge and node update functions. The superscript $l$ denotes the current message passing iteration with $n$ being the total number of message passing layers. In this context, geometric and molecular graphs used in this work differ by the definition of $N(i)$ and the edge feature $u_{ij}$.

\subsection{Energy Decomposition}\label{sec:intra}
Essential to the ANA2B model is a decomposition of interactions, which aims to follow a physically motivated classical description of interatomic interactions where possible. Remaining interactions are treated as corrections parametrized by ML models. The decomposition achieves two goals: First, the total potential energy is separated into manageable pieces. Second, the resulting interactions are interpretable. Here, a brief description of the involved interaction terms is given.
Based on the classical description assumed in this work, interactions are separated into purely intermolecular and purely intramolecular contributions, as well as dispersion interactions (D3),
\begin{equation}
    V_{\text{Total}} = V_{\text{Intra}} + V_{\text{Inter}} + V_{\text{D3}}.
\end{equation}
Dispersion interactions $V_{\text{D3}}$ are described with the D3 dispersion correction~\cite{D3} with Becke-Johnson damping with parameters for {PBE0}~\cite{D3BJ, BJ}, and are applied to both intramolecular and intermolecular interactions.

The purely intramolecular term $V_{\text{Intra}}$ is described in the ANA2B model by a ML potential, referred hereafter as $V_{\Delta\text{ML}}$. This ML potential was trained on energies and gradients of small molecules using PBE0/def2-TZVP as the reference method.

The purely intermolecular term $V_{\text{Inter}}$ consists of
\begin{equation}
    V_{\text{Inter}} = V_\text{ES} + V_{\text{Pol}} + V_{\text{$\Delta$SR}}.
\end{equation}
where $V_{\text{Pol}}$ refers to the polarization energy and $V_{\text{$\Delta$SR}}$ to the short-range two-body ML correction.
A detailed description of the intermolecular terms is given in the following paragraphs.

\subsubsection{Electrostatics}
Long-range intermolecular electrostatic interactions are described with atomic multipoles.
We made use of our previously introduced formalism for the prediction of atomic multipoles~\cite{MultipolesMe} based on MBIS atomic multipoles~\cite{MBIS} up to the quadrupole and atomic volumes. Here, the model is re-trained and improved based on the model architecture described in Ref.~\cite{AMP}.
Implementation of the electrostatic interaction and Ewald summation follows the formalism outlined in Refs.~\cite{MultipoleMethod, MultipoleMethodSmith, MultipoleMethodLin, EwaldMultipoles}.
The interaction of point multipoles at site $i$ and site $j$ is described as~\cite{MultipoleMethod},
\begin{equation}
     V_\text{ES}=\sum_{l=0}^4 B_l(r_{ij})G^l(\vec{r}_{ij}) .
\end{equation}
Note that intramolecular electrostatic interactions are contained in $V_{\Delta\text{ML}}$ (see above).

\subsubsection{Polarization}
A description of polarization is introduced through the Applequist model~\cite{Applequist} including Thole damping~\cite{Thole} as the energy resulting from placing the molecule in the electric field produced by the static multipoles,
\begin{equation}
    V_{\text{Pol}}=-\frac{1}{2}\mu_{\text{Ind}} E_{\text{Static}} ,
\end{equation}
where $\mu_{\text{Ind}}$ refers to the self-consistently converged induced dipoles, and $E_{\text{Static}}$ to the electric field produced by the static multipoles. $E_{\text{Static}}$ is not damped and includes only intermolecular contributions.
Induced dipoles $\mu$ are obtained as
\begin{equation}
     \mu_{\text{Ind}} = \textbf{B}^{-1}E_{\text{Static}}
\end{equation}
via inversion of the $3N\times 3N$ polarizability matrix $B$~\cite{Stone},
\begin{equation}
    \textbf{B} = \begin{cases}
                   \alpha^{-1}_{ij} & \text{for } i = j\\
                   -\textbf{T}_{ij} & \text{for } i \neq j\\
                    \end{cases}
    \label{eq:polarize_c6}
\end{equation}
with the atomic polarizability $\alpha_i$ and the elements $\textbf{B}_{ij}$ of the dipole-dipole interaction matrix.
These elements $\textbf{B}_{ij}$ are damped with the damping proposed by Thole,
\begin{equation}
    f_\text{Thole}=1-\exp(-au_{ij}^3)
\end{equation}
using a damping factor $a$ and the polarizability-normalized distance
\begin{equation}
    u_{ij}=\frac{r_{ij}}{(\alpha_i\alpha_j)^{\frac{1}{6}}} .
\end{equation}
The damping factor $a$ is set to $0.39$ as in the AMOEBA FF~\cite{AMOEBA}. 
For the first order polarization model ANA2B$^1$, $\mu_{\text{Ind}}$ was obtained as
\begin{equation}
     \mu_{i, \text{Ind}} = \alpha_i \cdot E_{i, \text{Static}} ,
\end{equation}
i.e., taking only the direct polarization into account. Thole damping is not applied to the direct polarization term.
Periodic boundary conditions are introduced through the Ewald summation formalism described in Ref.~\cite{EwaldPolarization}. The reciprocal space contribution is neglected for the mutual polarization term.

Static atomic dipole polarizabilities are obtained as
\begin{equation}
    \alpha_i = \alpha_0  \cdot \langle r_3 \rangle \cdot \phi_\alpha (h_i^2) ,
\end{equation}
where $\alpha_0$ is the polarizability of the isolated element, and $\langle r_3 \rangle$ the ratio between the atomic volume of the isolated atom and the atom in the molecule analogous to the Tkatchenko-Scheffler model~\cite{TSDispersion}.
Finally, an atom type derived scaling factor $\phi_\alpha (h_i^2)$ is introduced to calibrate the polarizabilities with respect to the dataset published in Ref.~\cite{AlphaML}. 
Atomic volumes$\langle r_3 \rangle$ are predicted by the same model that predicts the atomic multipoles, i.e., for the isolated molecule using MBIS atomic volumes~\cite{MBIS} as the reference.

\subsubsection{Short-Range Correction ($\Delta$SR)}\label{sec:short_range}
Instead of developing corrections for short-range phenomena such as charge penetration, a NN-parametrized pairwise interaction is proposed.  
This short-range pairwise potential is composed of the following terms,
\begin{equation}
    V_{\text{$\Delta$SR}} = V_{\text{Ex, Static}} + V_{\text{Ex, Ind}} + V_{\text{Att}},    
\end{equation}
and is applied to all intermolecular atom pairs within a distance of $6.5\,$\AA. 

The repulsive terms $V_{\text{Ex, Static}}$ and $V_{\text{Ex, Ind}}$ build on the orbital overlap model proposed by Salem~\cite{Salem} and extended by Murrell \textit{et al.}~\cite{S2R2}, which describes the exchange energy as a function of the orbital overlap $S^2$,
\begin{equation}
    V_{\text{Ex}} = \frac{K_{1} S_{}^2}{r_{}} + \frac{K_{2} S_{}^2}{r_{}^2} .
\end{equation}
Attractive contributions to the short-range interaction due to charge transfer and charge-penetration effects are introduced as
\begin{equation}
    V_{\text{Att}} = -K S^2.
\end{equation}
Parameters for these interaction terms (coupling parameters $K$ and overlaps $S^2$) are parametrized by a ML model. The input features are described in the following.

\begin{itemize}
\item \textit{Pairwise Atom Types:} Features obtained from molecular graphs $\mathcal{G_{\text{Mol}}}$ described in Section~\ref{sec:molecular_graphs} are symmetrized as $h_{ij}^1 = \phi_{h}(h_i^1, h_j^1) + \phi_{h}(h_j^1, h_i^1)$. For the short-range correction, only first-order atom types are used. These will be referred to as $h_{ij}^1$. Only first-order atom types are used to avoid overfitting as multipoles already include information about the environment.

\item \textit{Distance Features:} Distances are encoded with five Gaussians $\exp{(-\alpha r_{ij}^2)}$ with logarithmically spaced $\alpha \in {0.1, 1}$~\AA$^{-2}$. 
Preliminary work (data not shown) indicated that Bessel functions, which are frequently used to encode distances, induce oscillations in the pairwise potential. The Gaussians are centered at $0$ to avoid this behaviour. These features will be referred to as $d_{ij}$.

\item \textit{Anisotropic Features:}
Anisotropy is introduced based on the atomic multipoles $M^k$ of order $k$ as the symmetrized multipole-multipole interaction coefficients~\cite{MultipoleMethod}, 
\begin{equation}
    \begin{aligned}
        g_0 &= M_i^0 \cdot M_j^0\\
        g_1 &= M_i^0 \cdot (M_{j, \alpha}^1  \vec{r}_{ij, \alpha}) - M_j^0 \cdot (M_{i, \alpha}^1  \vec{r}_{ij, \alpha})\\
        g_2 &= (M_{i, \alpha}^1  \vec{r}_{ij, \alpha}) \cdot (M_{j, \alpha}^1  \vec{r}_{ij, \alpha})\\
        g_3 &= M_{i, \alpha}^1 M_{j, \alpha}^1\\
        g_4 &= (M_{i, \alpha\beta}^2 \vec{r}_{ij, \alpha})_\beta M_{j, \beta}^1 - (M_{j, \alpha\beta}^2  \vec{r}_{ij, \alpha})_\beta M_{i, \beta}^1\\
        g_5 &= M_j^0 \cdot (M_{i, \alpha\beta}^2 \vec{r}_{ij, \alpha\beta}) + M_i^0 \cdot (M_{j, \alpha\beta}^2 \vec{r}_{ij, \alpha})\\
        g_6 &= M_{i, \alpha\beta}^2 M_{j, \alpha\beta}^2 \\
        g_7 &= (M_{i, \alpha\beta}^2 \vec{r}_{ij, \alpha})_\beta (M_{j, \alpha\beta}^2 \vec{r}_{ij, \alpha})_\beta \\
        g_8 &= (M_{j, \alpha\beta}^2 \vec{r}_{ij, \alpha\beta}) \cdot (M_{i, \alpha}^1 \vec{r}_{ij, \alpha}) -(M_{i, \alpha\beta}^2 \vec{r}_{ij, \alpha\beta}) \cdot (M_{j, \alpha}^1 \vec{r}_{ij, \alpha})\\        
        g_9 &= (M_{i, \alpha\beta}^2 \vec{r}_{ij, \alpha\beta}) \cdot (M_{j, \alpha\beta}^2 \vec{r}_{ij, \alpha\beta}) .
    \end{aligned}
\end{equation}
In this context, scalar multiplication is indicated by $\cdot$ and contractions are performed over the Cartesian components indicated by the greek indices. $\vec{r}_{ij, \alpha\beta}$ refers to the tensor product of the Euclidean vector $\vec{r}_{ij, \alpha} = \vec{r}_j - \vec{r}_i$ with itself. Vectors $\vec{r}_{ij, \alpha}$ are normalized.
Two types of features are used. The first type is calculated without inclusion of the induced dipoles, i.e., $(M^0, M^1, M^2)$, whereas the second includes the contribution of the induced dipoles, i.e., $(M^0, M^1 + \mu_{\text{Ind}}, M^2)$. These features will be referred to as $g_{ij, {\text{Static}}}$, and $g_{ij, {\text{Ind}}}$, respectively.
\end{itemize}

Using the above features, the orbital overlaps $S^2$ are parametrized by an ANN $\phi_{S^2}$ as,
\begin{equation}
    \begin{aligned}
        S_{\text{Att}}^2 &= \phi_{S^2, \text{Static}}(h_{ij}^1, d_{ij}, g_{ij, {\text{Static}}})\\   
        S_{\text{Ex, Ind}}^2 &= \phi_{S^2, \text{Ind}}(h_{ij}^1, d_{ij}, g_{ij, \text{Ind}})\\
        S_{\text{Ex, Static}}^2 &= \phi_{S^2, \text{Static}}(h_{ij}^1, d_{ij}, g_{ij, {\text{Static}}})\\
    \end{aligned}
\end{equation}
Overlaps sharing the same input features, i.e., $S_{\text{Att}}^2$ and $S_{\text{Ex, Static}}^2,$ are predicted by the same model. 

Coupling constants $K$ are predicted as
\begin{equation}
    K = \phi_{K}(h_{ij}^1, d_{ij}),
\end{equation}
that is without including the anisotropic features. Independent coupling constants are predicted for each term using the same model. Overlaps and Gaussian distance features are multiplied with a switching function to guarantee a smooth cutoff~\cite{SwitchingFunction},
\begin{equation}
    \begin{aligned}
    f_{\text{Switch}}(x) &= 1 - 6x^5 + 15x^4 - 10x^3\\
    x(r) &= \frac{(r - r_{\text{switch}})}{(r_{\text{cut}} - r_{\text{switch})}}
    \end{aligned}
\end{equation}
with distance $r$, cutoff $r_{\text{cut}}$, and switching distance $r_{\text{switch}}$. The switching distance is set to $r_{\text{cut}}-1$\,\AA.


\section{Methods}

\subsection{Models and Training Procedure}\label{sec:dataset_intra}
\begin{figure}[H]
    \centering
    \includegraphics[width=0.8\textwidth]{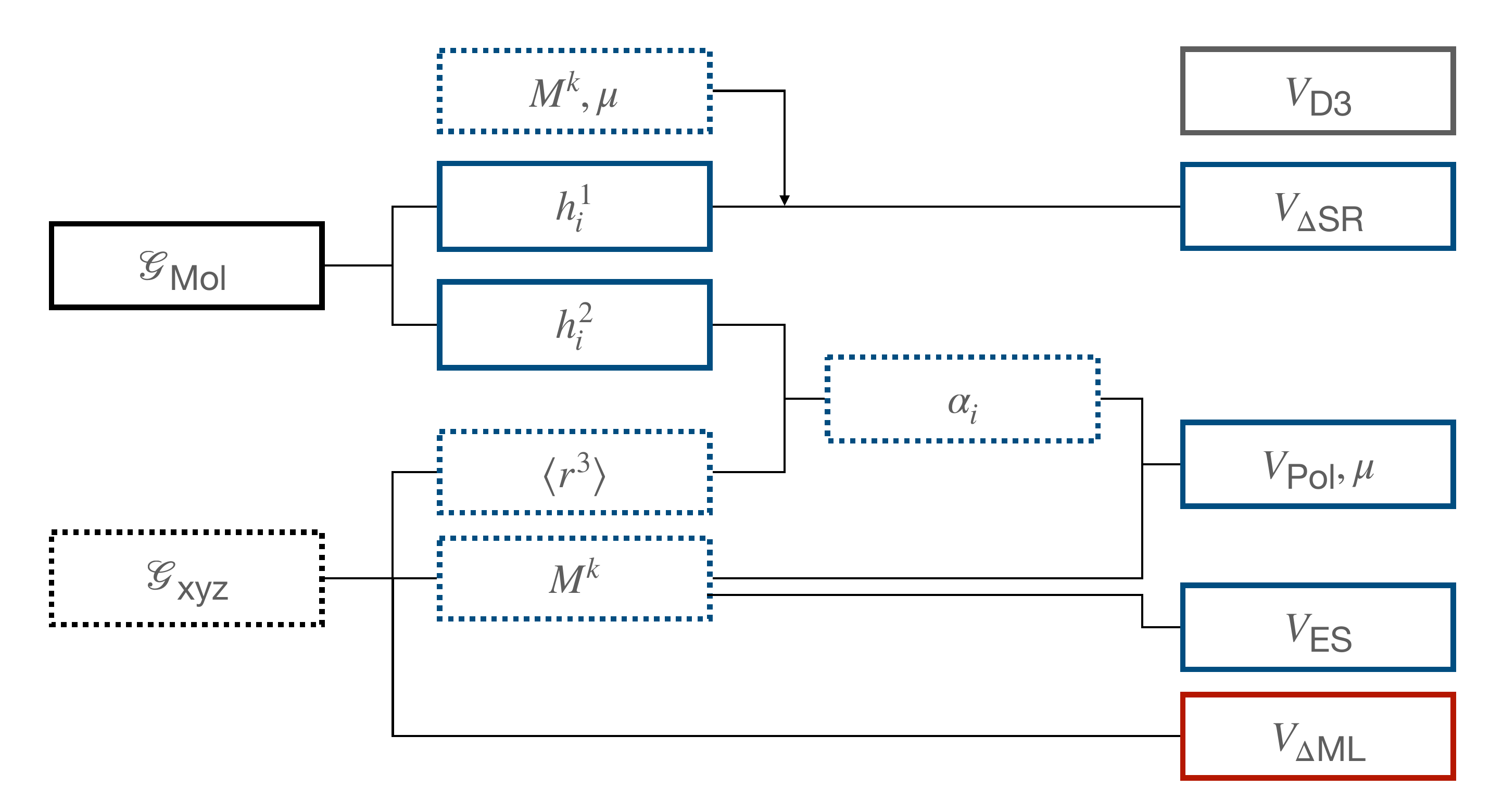}
    \caption{Overview of the ANA2B model. Dotted lines refer to features that depend on the geometry while bold lines to features based on molecular graphs. Blue components refer to intermolecular interactions, red to intramolecular interactions, and grey to shared interactions.}
\label{fig:overview}
\end{figure}

Several ML models were used in this work. An overview is given in Figure \ref{fig:overview}. If not mentioned otherwise, ANN parametrized functions $\phi$ were constructed from two fully connected feed-forward layers of size $128$ using the Swish activation function~\cite{Swish}. The GNNs used to extract features of molecular graphs $\mathcal{G_{\text{Mol}}}$ used a node-embedding and edge-embedding layer and message passing layers consisting of a single feed-forward layer of size $64$. Each model was trained separately on its respective target. If not noted otherwise, models were optimized with Adam~\cite{ADAM} using an exponentially decaying learning rate $\in [5\cdot 10^{-4}, 1 \cdot 10^{-5}]$.

\subsubsection{Multipoles and Atomic Volumes}
MBIS multipoles on a {PBE0}/def2-TZVP level of theory were predicted using our previous formalism for an equivariant multipole GNN~\cite{MultipolesMe}. In addition, the MBIS atomic volume ratio was included. The message passing formalism described in Ref.~\cite{MultipolesMe} was replaced with the AMP formalism described in Ref.~\cite{AMP}. For training, the dataset generated in Ref.~\cite{MultipolesMe} was used and extended with conformations sampled with molecular dynamics (MD) to improve coverage of off-equilibrium conformations. MD simulations were performed with xTB (version 6.4.1)~\cite{XTB} using the GFN-1 Hamiltonian~\cite{GFN1}. A seed conformation for MD was generated with the ETKDG conformation generator~\cite{ETKDG} as implemented in the RDKit~\cite{RDKIT}. MD simulations were carried out in the NVT ensemble for $n \cdot 100\,$ps, with integration steps of $0.5\,$fs at $800\,$K without any constraints. If not stated otherwise, default settings (sccacc$=2$, hmass$=4\,$a.u.) were used. $n$ was determined based on the number of heavy atoms in the molecule ($<$5:~$n=16$, $<$7: $n=8$, $<$11: $n=4$, $>$10: $n=2$). Snapshots were written out every $100\,$ps. The $n+1$ conformations, including the xTB GFN-1 minimum structure, obtained in this manner served as input for the following single-point calculations. Single-point gradients were evaluated for each structure with {PBE0}/def2-TZVP~\cite{PBE, PBE0, DEF2} using PSI4 (version 1.4)~\cite{PSI4, PSI4_1.1}. MBIS multipoles~\cite{MBIS} and volumes were obtained with PSI4~\cite{PSI4_1.4}. If not stated otherwise, default PSI4 settings were used (energy and density convergence threshold $10^{-8}\,$a.u.). Data for $1'514'462$ conformations for a total of $451'973$
unique molecules were obtained in this way.

\subsubsection{ML Correction}
The ML correction was used to describe intramolecular interactions except for the contribution of the D3 dispersion model. 
The $\Delta$ML potential is based on the AMP architecture proposed in Ref.~\cite{AMP}. However, instead of a single set of multipoles, a total of $32$ independent sets of multipoles up to the quadrupole were expanded on each atom. Note that these multipoles serve only as a tool to introduce directional interactions, unlike the electrostatic multipoles used for $V_\text{ES}$. Three message passing steps were employed with a cutoff of $5\,$\AA.
The model was trained on the difference between the reference {PBE0} potential energy and gradient as well as the potential energy and gradient of the bond-stretching and damped electrostatic interactions. The model was trained on the same dataset used to train the multipole model. The model was trained over $2'048$ epochs. Gradient norms were clipped to norm $1$. During each epoch, $1'028$ samples were presented. Each sample consisted of a batch of all conformations of five molecules. The model was trained on weighted relative energies and gradients,
\begin{equation}
    \mathcal{L}_{\Delta \text{ML}}= w_i \cdot (1 - \beta) \cdot 
    (\Delta V_{\text{Ref}} - \Delta V_{\text{ML}})^2 + \frac{\beta}{3 N}\sum_i^N \sum_\alpha^3 \left(\frac{\partial V_{\text{Ref}}}{\partial x_{i, \alpha}} - \frac{\partial V_{\text{ML}}}{\partial x_{i, \alpha}}\right)^2.
\end{equation}
$\Delta V_{\text{ML}}$ and $\Delta V_{\text{Ref}}$ refer to the relative energies, i.e., the difference between the energy of a conformation $i$ and a conformation $j$ serving as a reference point $\Delta V = V_i - V_j$.  $\beta$ was set to $0.9$. Weights $w_i$ were defined as, 
\begin{equation}
    w _i = \exp\left (-\frac{V_{min} - V_{i}}{k_b T N}\right ) ,
\end{equation}
where $V_{min}$ is the energy of the conformation with the lowest energy of a given molecule, and $N$ to the numbers of atoms.
$T$ was set to $2'000\,$K. Only molecules with more than one possible conformation and conformations with negative atomization energies and with maximum gradient components $\leq 2'000\,$kJ/mol\AA~were used.

\subsubsection{Short-Range Correction}
\label{sec:training_sr}
The short-range pairwise potential $V_{\Delta \text{SR}}$ was trained on the intermolecular potentials of dimers in vacuum from the DES5M dataset~\cite{DEShawDimers}. A cutoff of $6.5$\,\AA~ was used for this interaction. As an exception, a cutoff of $5.5$\,\AA~was found to be optimal for the ANA2B$^0$ model, i.e., the model without any polarization interactions.
The model was trained over $512$ epochs. Gradient norms were clipped to norm $1$. During each epoch, $2'048$ samples were presented. Each sample consisted of all configurations of a given dimer. The mean squared error (MSE) between the predicted intermolecular potential and the reference (SNS-MP2)~\cite{SNS, SNSMP2} was optimized. Performance on the S7L~\cite{S7L, QCMRef} and S66x8~\cite{S66x8} datasets were used as signals for early stopping. The mean absolute error (MAE) on a set of structures from X23 and ICE13 (CYTSIN01, URACIL, UREAXX12, HXMTAM10, CYHEXO, SUCACB03, CYANAM01, PYRZOL05, OXALAC04, ammonia, CO2 and ice polymorphs Ih and II) was used to the select the final models.

\subsubsection{Polarizabilities}
The model used to predict polarizability scaling factors from molecular graphs was trained on a dataset of CCSD molecular polarizabilities reported in Ref.~\cite{AlphaML}. The model was
trained over $512$ epochs. During each epoch $512$ randomly drawn samples consisting of a single molecule were presented. The model was optimized with respect to the MSE between the predicted molecular polarizability and the CCSD molecular polarizability.

\subsection{General Implementation Details}
All ML models were implemented in TensorFlow (version 2.11.0)~\cite{TensorflowPaper}.
The atomic simulation environment (ASE, version 3.22.1)~\cite{ASE} was used as MD engine, for optimization, and for general analysis tasks including the calculation of harmonic free energies and thermodynamic integration. 
MDTraj (version 1.9.8)~\cite{MDTraj} was used for post-processing and analysis tasks.

For long-range electrostatic interactions and polarization, a real-space cutoff of $10\,$\AA~was used. 
The screening parameter $\alpha$ for Ewald summations was set to $0.292$ and $0.215$ for the evaluation of the electrostatic interaction and the mutual polarization, respectively. Crystal structures were minimized with fixed lattice parameters. For MD simulations involving liquids, cutoffs for the D3 model were set to 
$10\,$\AA, $5\,$\AA, and $10\,$\AA~for the two-body-term, three-body-term, and the coordination number, respectively.
For calculations and MD simulations involving crystals, cutoffs for the D3 model were set to $15\,$\AA, $8\,$\AA, and $15\,$\AA~for the two-body-term, three-body-term, and the coordination number, respectively.

\subsection{Molecular Dynamics (MD) Set-up}
Simulations of the pure liquids in the GROMOS 2016H66 dataset were performed with ASE~\cite{ASE}. $22$\,\AA~cubix boxes were generated with packmol~\cite{Packmol} followed by a pre-equilibration over $10'000$ steps at $300$\,K with OpenFF (version 2.0) using OpenMM (version 8.0)~\cite{OpenFF2, OpenMM7}.
Equilibration and production runs were performed with an Andersen thermostat~\cite{AndersenThermostat} at the simulation temperature described in the GROMOS 2016H66 publication~\cite{2016H66} ($298.15$\,K if not noted otherwise) and a Monte-Carlo barostat~\cite{MCBarostat} with a target presssure of $1$\, bar.
The integration step was set to $0.5$\,fs.
The equilibration was performed over $2'000$ steps ($1$\,ps) using the respective ANA2B model with the collision frequency set to $0.1$ and the barostat frequency set to $10$.
For the production run over $100'000$ steps ($50$\,ps), the collision frequency was set to $0.01$ and the barostat was applied every $25$th step. These runs were repeated three times with different random number seeds for the generation of the initial velocities. Ensemble properties were averaged over the last $25$\,ps.

For the prediction of the heat of vaporization, monomers were simulated in the gas phase. These simulations were equilibrated over $2'000$ steps ($1$\,ps) using a Berendsen thermostat~\cite{BerendsenThermostat} ($\tau=10$\,fs) followed by a $100'000$ step ($50$\,ps) production run using a Langevin thermostat with a friction of $1$\,a.u. Starting conformations were generated with the ETKDG conformation generator~\cite{ETKDG} in the RDKit~\cite{RDKIT}. Again, averages were taken over four replicates with different initial velocities.

\subsection{Ranking of Crystal Structures -- CSP Blind Tests 3 and 5}\label{sec:blind_tests15}
For the third CSP blind test~\cite{CSP3BT}, all structures submitted by van Eijck were used (entries VIII, X, XI)~\cite{vanEijck0, vanEijck1, vanEijck2}. For the fifth blind test~\cite{CSP5BT}, submissions of Neumann and co-workers were considered (entries XVI, XVII, XVIII)~\cite{NeumannCSP, TFF1, TFF2}. These submissions were selected because they contain in all cases a candidate structure that was considered a match with the experimental structure.

Candidate structures were relaxed under fixed lattices using the (L)BFGS optimizer with a tolerance of $1$\,kJ/mol\AA~\cite{BFGS1, BFGS2, BFGS3, BFGS4, LBFGS}. Lattices parameters were not minimized.
Structures that did not converge within $250$ steps were excluded.

\subsection{Ranking of Crystal Structures -- CSP Blind Test 6}
\subsubsection{Relaxtion of Crystal Structures} 
Lattices were relaxed with an external pressure of $1$\,bar using an anisotropic Monte Carlo barostat~\cite{MCBarostat} at $0$\,K.
Subsequently, structures were relaxed under fixed lattices using the (L)BFGS optimizer with a tolerance of $1$\,kJ/mol\AA~\cite{BFGS1, BFGS2, BFGS3, BFGS4, LBFGS}. 

\subsubsection{MD Simulations} 
The NPT ensemble was sampled using an Andersen thermostat~\cite{AndersenThermostat} and an anisotropic Monte Carlo barostat~\cite{MCBarostat} at $1$\,bar and a temperature of $150$\,K (XXII) and $300$\, (XXIII, XXVI). The collision frequency was set to $0.1$ and the barostat frequency was set to $10$. Structures were equilibrated for $1$\,ps followed by $5$\,ps production runs. These simulations were used to obtain thermally expanded cells and the mean potential energy.

\subsubsection{Gibbs Term} 
The difference between the Helmholtz free energy $F$ and the Gibbs free energy ($G$) was obtained as ~\cite{Gibbs},
\begin{equation}
    \Delta_{F\rightarrow G} = P\langle V \rangle + k_BT\log \rho(h|P, T),
\end{equation}
with $\langle V \rangle$ referring to the mean volume during the simulation and P to the pressure. The density $\rho(h|P, T)$ was obtained through a kernel density estimation using Gaussian kernels with a width of $0.1$. The density was estimated for the cell parameters.

\subsubsection{Helmholtz Free Energy} 
The harmonic Helmholtz free energy ($F_H$) was calculated with the phonon module implemented in ASE using the minimized structures from step one. The phonon density of states was sampled on a uniform $k$-point grid of size ($20$, $20$, $20$) using $2'000$ sampling points.

\subsubsection{Thermodynamic Integration} 
The anharmonic correction to the harmonic Helmholtz free energy ($F_A$) was obtained with a thermodynamic integration from the harmonic potential ($V_H$) to the unconstrained potential ($V_A$), 
\begin{equation}
    \Delta _{H\rightarrow A} = \int_{0}^{1} \langle V_A - V_H\rangle_\lambda d\lambda
\end{equation}
following the description in Ref.~\cite{FECrystal}. The harmonic potential was obtained from the numerically calculated Hessian of the relaxed structure using the lattice parameters with the highest likelihood. The thermodynamic integration was performed over eleven uniformly spaced $\lambda$-points. 
Numerical integration was performed using a trapezoidal integration. An initial equilibration over $500$\,fs was performed followed by $100$\,fs of equilibration and $1$\,ps of sampling at each lambda point. The NVT ensemble was sampled with an Andersen thermostat~\cite{AndersenThermostat} at $150$\,K (XXII) and $300$\,K (XXIII, XXVI). 


\section{Results and Discussion}

The proposed ANA2B model was applied to a range of existing benchmarks to establish a level of accuracy. The datasets are categorized by their use as training, validation, or test set, and include intermolecular potential energies of dimers and lattice energies of molecular crystals and water ice.
Particular attention is given to the role of polarization because preliminary results (data not shown) highlighted its importance. We have thus studied three variations of the ANA2B model: The first variation does not include any polarization interaction at all, and will be referred to as ANA2B$^0$. The second variation, labelled ANA2B$^1$, includes only the polarization stemming from the direct field, i.e., neglecting the mutual polarization. The third variation, labelled ANA2B$^\infty$, includes a full treatment of the direct and mutual polarization terms. At present, all models were only trained and applied to neutral molecules consisting of the elements H, C, N, O, F, S, Cl.

\subsection{Monomers in Vacuum}
\subsubsection{Performance on Training and Validation Sets}
A dataset of small molecules, covering potential energies, gradient, atomic multipoles, and atomic volume ratios on a PBE0/def2-TZVP level of theory was used to train the intramolecular potential. The construction of this dataset is discussed in Section~\ref{sec:dataset_intra}. Table \ref{tab:intra_results_train} reports the errors for the gradients and relative energies for the training set and validation set.

\begin{table}[H]
    \caption{Mean absolute error (MAE) in [kJ/mol] for the training set and validation set of energies and gradients of small molecules in vacuum. Errors for relative energies, i.e., with respect to the energy of a reference conformation, are reported. Three outlier conformations were excluded due to highly deformed structures being present. More detailed error statistics is provided in Table S1 in the Supporting Information.} 
    \centering
    \small
    \begin{tabular}{l c c | c}
    \hline
    Name & N  & Type &  $\Delta$ML$_\text{Intra}$   \\\hline
    $\Delta$Energy & 1'398'301 & Train & 0.5\\
    Gradient &   & Train & 0.8\\
    $\Delta$Energy & 79'369 & Val & 0.6\\
    Gradient &   & Validation & 0.8\\\hline
    \end{tabular}    
    \label{tab:intra_results_train}
\end{table}

\subsubsection{Performance on Test Set}
The following section reports the performance of the intramolecular ML potential on several computational benchmark datasets of conformation energies.
Overall, we find that our model performs comparable to the reference method (PBE0-D3BJ/def2-TZVP) with MAE values that are typically larger by a few tenths of a kJ/mol. 
These results justify on one hand the decision to use the ML potential in place of the DFT calculation and that ML potentials might overall be able to substitute DFT in many situations. At the same time, datasets such as PCONF clearly display how the ML potential `inherits' the accuracy of the method used to generate the training set.

\begin{table}[H]
    \caption{Mean absolute error (MAE) in [kJ/mol] for the intramolecular ML potential used in this work for benchmarks of conformation energies (test sets). $N$ is the number of data points per dataset. The method used to generate the training data (PBE0-D3) is shown as a comparison. More detailed error statistics is provided in Table S2 in the Supporting Information.} 
    \centering
    \small
    \begin{tabular}{l c c | c c c c c c c}
    \hline
    Name & $N$ & Type & Intra ML & PBE0-D3\\\hline
    Glucose~\cite{CarbohydratesBenchmark} & 205 & Test & 2.5 & 2.3\\
    Maltose~\cite{CarbohydratesBenchmark} & 223 & Test & 2.7 & 1.9\\
    SCONF~\cite{SCONF} & 17 & Test & 1.6 & 1.1\\
    PCONF~\cite{PCONF} & 10 & Test & 6.7 & 6.2\\
    ACONF~\cite{ACONF} & 15 & Test & 0.5 & 0.2\\
    CYCONF~\cite{CYCONF} & 15 & Test & 2.7 & 2.7\\\hline
    \end{tabular}
    
    \label{tab:monomer_results_test}
\end{table}

\subsection{Dimers in Vacuum}
\subsubsection{Performance on Training Set}
Table~\ref{tab:dimer_results_train} displays MAEs for the full training set (DES5M). In all cases, the prediction error of around $2.0\,$kJ/mol is below the `chemical accuracy' level of $4.184\,$kJ/mol. If only near-equilibrium structures ($<$10~kJ/mol) are considered, the MAE drops further to $0.5\,$kJ/mol. For a subset of 370'000 molcules (DES370K), CBS extrapolated CCSD(T) reference data exists, which was used to train the SNS-MP2 model~\cite{DEShawDimers} applied to the remaining DES5M dataset. Compared to SNS-MP2 itself ($0.2\,$kJ/mol for DES370K and $0.1\,$kJ/mol for DES370K$_{<\text{10kJ/mol}}$~\cite{DEShawDimers}), the ANA2B$^{\infty}$ model introduces an additional error of $0.9\,$kJ/mol. On near-equilibrium structures (DES370K$_{<\text{10kJ/mol}}$), our model introduces only an additional $0.4\,$kJ/mol error compared to the error between SNS-MP2 and CCSD(T)/CBS.

\begin{table}[H]
    \caption{Mean absolute errors (MAE) in [kJ/mol] for the training set of DES5M (SNS-MP2) and the subset DES370K (CCSD(T)/CBS). $^1$ For the DES370K subset, MAE values with respect to the CCSD(T)/CBS reference are reported. The models were trained on the full SNS-MP2 dataset (DES5M). $N$ is the number of data points per dataset. More detailed error statistics is provided in Table S3 in the Supporting Information.} 
    \centering
    \small
    \begin{tabular}{l c c | c c c c c c c c}
    \hline
    Name & $N$  & Type & ANA2B$^0$ & ANA2B$^1$ & ANA2B$^{\infty}$  \\\hline
    DES5M~\cite{DEShawDimers} & 4'034'267 & Train & 1.9 & 2.0 & 2.0\\
    DES5M$_{<\text{10kJ/mol}}\,$\cite{DEShawDimers} & 3'255'535 & Train & 0.5 & 0.5 & 0.5\\
    DES370K~\cite{DEShawDimers} & 269'611 & Train$^1$ & 1.2 & 1.2 & 1.1\\
    DES370K$_{<\text{10kJ/mol}}\,$\cite{DEShawDimers} & 235'958 & Train$^1$ & 0.5 & 0.5 & 0.5\\\hline
    \end{tabular}
    \label{tab:dimer_results_train}
\end{table}

\subsubsection{Performance on Validation Set}
The S66x8~\cite{S66x8} and S7L~\cite{S7L} datasets were used as early-stopping signal during training of the ANA2B models. While only small differences are found for the small molecule dimers in the S66x8 dataset, a considerably larger MAE is observed for the supramolecular systems in the S7L dataset. These results are consistent with the results observed for molecular systems shown below in Subsection~\ref{sec:results_late}. Very large molecules and/or molecular clusters might thus be an adequate and cost-efficient substitute to train and validate size-transferable ML potentials in the absence of condensed-phase data.
For the S7L structures, the PNO coupled cluster calculations of Ref.~\cite{QCMRef} were used.

\begin{table}[H]
    \caption{Mean absolute error (MAE) in [kJ/mol] for the validation sets S66x8~\cite{S66x8} and S7L~\cite{S7L, QCMRef}. $N$ is the number of data points per dataset. More detailed error statistics is provided in Table S4 in the Supporting Information.} 
    \centering
    \small
    \begin{tabular}{l c c | c c c}
    \hline
    Name & N & Type & ANA2B$^0$ & ANA2B$^1$ & ANA2B$^{\infty}$\\\hline
    S66x8~\cite{S66x8} & 528 & Validation & 1.3 & 0.8 & 0.8 \\
    S7L~\cite{S7L, QCMRef} & 7 & Validation & 21.1 & 2.0 & 2.3\\\hline
    \end{tabular}    
    \label{tab:dimer_results_val}
\end{table}

\subsubsection{Performance on Test Set}
Table~\ref{tab:dimer_results_test} lists MAE values for 14 computational benchmark datasets of dimer interaction potentials.
In most cases, errors for the three ANA2B models are comparable. However, for datasets that contain highly polarizable systems, e.g., nucleobases in JSCH and ACHC, or for datasets with hydrogen-bonded systems, i.e., HB375x10, HB300SPXx10 and HBC1, the two models which include a treatment of polarization (ANA2B$^1$ and ANA2B$^{\infty}$) perform better.

\begin{table}[H]
    \caption{Mean absolute error (MAE) in [kJ/mol] for intermolecular potential benchmarks of dimers in vacuum (test sets). $N$ is the number of data points per dataset. More detailed error statistics are provided in Table S5-S7 in the Supporting Information.} 
    \centering
    \small
    \begin{tabular}{l c c | c c c c c c c c}
    \hline
    Name & N & Type & ANA2B$^0$ & ANA2B$^1$ & ANA2B$^{\infty}$\\\hline
    SSI~\cite{BioFragmentDB} & 2'596 & Test & 0.6 & 0.7 & 0.6\\
    BBI~\cite{BioFragmentDB} & 100 & Test & 1.0 & 0.7 & 0.7\\
    UBQ~\cite{UBQ} & 81 & Test & 1.0 & 1.2 & 1.0\\    
    ACHC~\cite{ACHC} & 54 & Test & 4.8 & 2.2 & 1.0\\
    JSCH~\cite{JSCH} & 123 & Test & 4.7 & 2.9 & 2.8\\
    HSG~\cite{HSG} & 16 & Test & 0.8 & 0.7 & 0.8\\
    HBC1~\cite{HBC} & 58 & Test & 8.5 & 3.3 & 2.0\\    
    S22~\cite{JSCH} & 22 & Test & 3.3 & 1.7 & 1.6\\
    S22x7~\cite{S22x7} & 154 & Test & 5.9 & 3.0 & 2.8\\
    D1200~\cite{NCIDispersion} & 482 & Test & 1.7 & 1.2 & 1.2\\
    D442x10~\cite{NCIDispersion} & 1'570 & Test & 1.9 & 1.6 & 1.5\\
    R739x5~\cite{NCIRepulsion} & 1'615 & Test & 2.5 & 2.3 & 2.3\\
    HB375x10~\cite{rezac2020b} & 3'750 & Test & 1.9 & 1.4 & 1.4\\
    HB300SPXx10~\cite{rezac2020a} & 1'210 & Test & 4.1 & 3.1 & 3.5\\\hline   
    \end{tabular}    
    \label{tab:dimer_results_test}
\end{table}

\subsection{Molecular Crystals}\label{sec:results_late}

\begin{figure}[H]
    \centering
    \includegraphics[width=\textwidth]{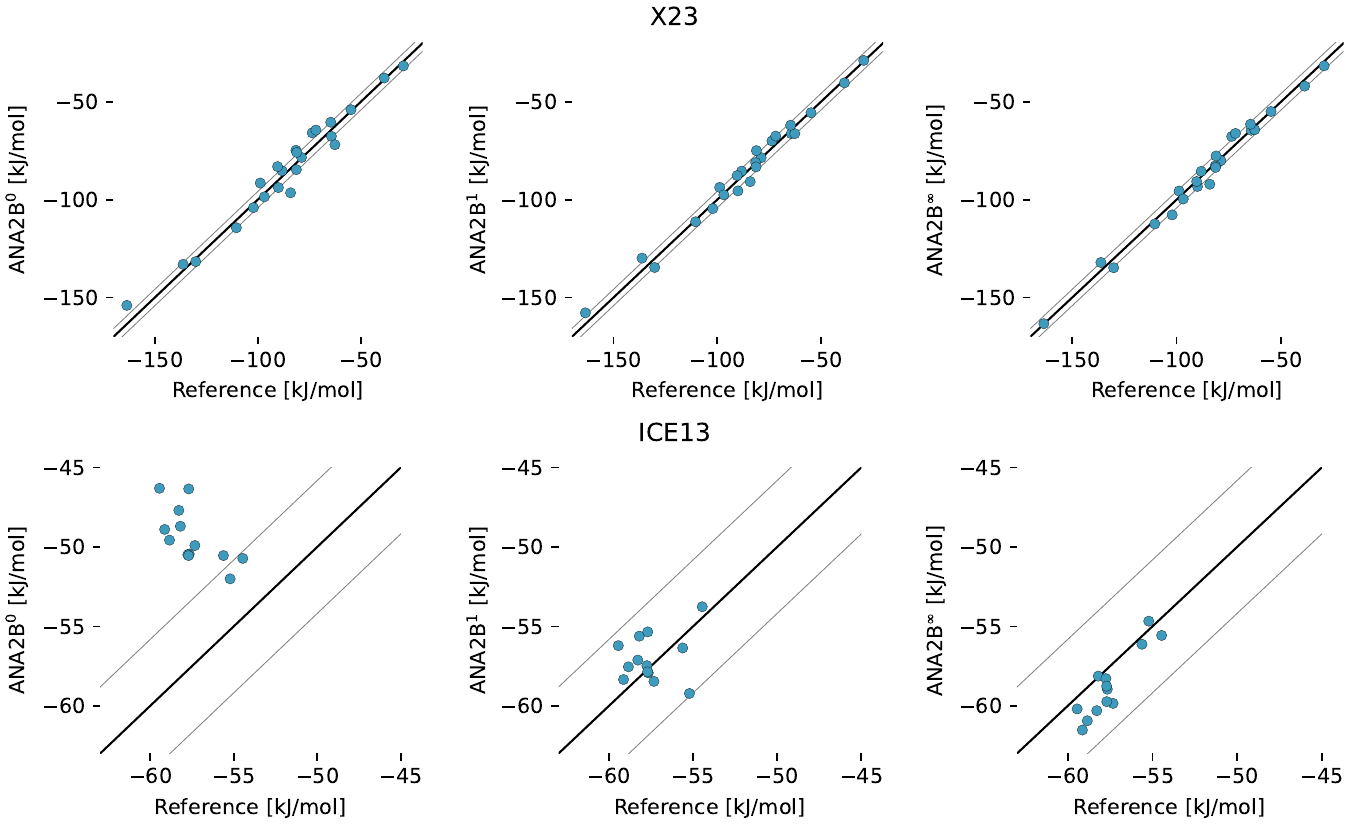}
    \caption{Results for the lattice energies of systems in the X23 (top panels) and ICE13 datasets (bottom panels) with the ANA2B models. The left column (ANA2B$^0$) refers to the model without any treatment of polarization, the middle (ANA2B$^1$) column shows results for the model that includes only the direct polarization term, and the right column (ANA2B$^\infty$) displays results for the model which includes a full treatment of polarization.
    Parity $\pm$\,4.184~kJ/mol is indicated by the black lines.}
    \label{fig:results_x23_ice}
\end{figure}

To assess whether the ANA2B can transfer from dimers and monomers in vacuum to condensed-phase systems, the model was applied to the prediction of lattice energies of molecular crystals.
Table~\ref{tab:cp_results_test} and Figure \ref{fig:results_x23_ice} show the corrected experimental lattice energies of the X23 dataset~\cite{X23Revised, X23Johnson, X23Tkatchenko} and diffusion Monte Carlo lattice energies for water ice polymorphs~\cite{ICE13}. Note that a subset of structures from X23 and ICE13 were used as validation structures to select the final model (see Sec.~\ref{sec:training_sr}).
Overall, the observed MAE is comparable to the most accurate dispersion corrected DFT calculations reported so far. For example, a recent study by Price \textit{et al.}~\cite{X23_XDM} reported an MAE of $2.0\,$kJ/mol using B86bPBE-25 in combination with the XDM dispersion correction. The same study also reported an MAE of $0.8\,$kJ/mol for the ICE13 dataset. Note that direct comparison with B86bPBE-25 is somewhat complicated by the fact that the lattice energies were obtained for structures minimized with a different method (B86bPBE). Finally, the MAE for the X23 dataset with an existing multipole FF for molecular crystals, FIT~\cite{FIT}, is reported as $9.2$\,kJ/mol~\cite{X23FF}. This direct comparison indicates that the hybrid approach proposed in this work may present a way to unlock the full potential of classical FFs.

The overall good performance of ANA2B compared to hybrid DFT methods is particularly interesting considering that hybrid DFT calculations are currently probably the most accurate approach feasible for relatively large scale studies of condensed-phase systems. Taking into account the error of the reference method itself and the error resulting from the ML model underscores the importance of developing ML models, which are transferable and thus able to take advantage of the high-quality data available for small systems.

In the case of the ice polymorphs, the importance of a description of polarization becomes evident. While the expensive treatment of mutual polarization (ANA2B$^{\infty}$) results only in a small improvement of the MAE compared to the (ANA2B$^1$), a clear difference is observed with regards to the ranking of the ice polymorphs (Table~\ref{tab:cp_results_test}): For the ANA2B$^{\infty}$ model, good agreement with the DMC reference is found with a Spearman correlation coefficient $r_\text{Spearman}$ of $0.77$. For the ANA2B$^1$, the ranking is considerably worse with a slightly negative coefficient $r_\text{Spearman} = -0.04$ (ANA2B$^0$: $r_\text{Spearman} = -0.74$).
While water presents a unique case, which might exaggerate the importance of polarization, these results still show a clear trend.
Including some description of the non-additive nature of polarization might thus be the most important ingredient required to achieve transferability from vacuum to the condensed phase.

\begin{table}[H]
    \caption{Mean absolute error (MAE) in [kJ/mol] for experimental (X23) and computationally (ICE13) derived lattice energies of molecular crystals in the test set. $N$ is the number of data points per dataset. Results for B86bPBE and B86bPBE-25 with XDM dispersion correction are taken from Ref.~\cite{X23_XDM}. B86bPBE-25 values were calculated with geometries relaxed at the B86bPBE level (B86bPBE-25//B86bPBE). $^\text{1}$Errors on a subset of X23 (cytosine, uracil, urea, hexamethylenetetramine, cyclohexane-1,4-dione, succinic acid, cyanamide, pyrazole, ammonia, and CO$_2$) and ICE13 (ice Ih, ice II) were used to select the final model. Graphical results are shown in Figure~\ref{fig:results_x23_ice}. More detailed error statistics are provided in Table S8-S10 in the Supporting Information.} 
    \centering
    \small
    \begin{tabular}{l c c | c c c c c c c c }
    \hline
    Name & N & Type & ANA2B$^0$ & ANA2B$^1$ & ANA2B$^{\infty}$ & B86bPBE & B86bPBE-25 \\\hline
    X23~\cite{X23Revised, X23Johnson, X23Tkatchenko} & 23 & Test$^\text{1}$ & 4.6 & 3.2 & 2.9 & 3.0 & 2.0\\
    DMC-ICE13~\cite{ICE13} & 13 & Test$^\text{1}$ &  8.3 & 1.4 & 1.3 & 7.5 & 0.8\\\hline
    \end{tabular}    
    \label{tab:cp_results_test}
\end{table}

\subsection{Condensed-Phase Properties of Pure Liquids}
Reproduction of experimental condensed-phase properties of molecular liquids have been a long-standing goal for the parametrization and testing of classical FF. Particularly for ML-based FF, these properties are an interesting test case as they require sufficient sampling in both the gas phase and the condensed phase.

\begin{figure}[H]
    \centering
    \includegraphics[width=1\textwidth]{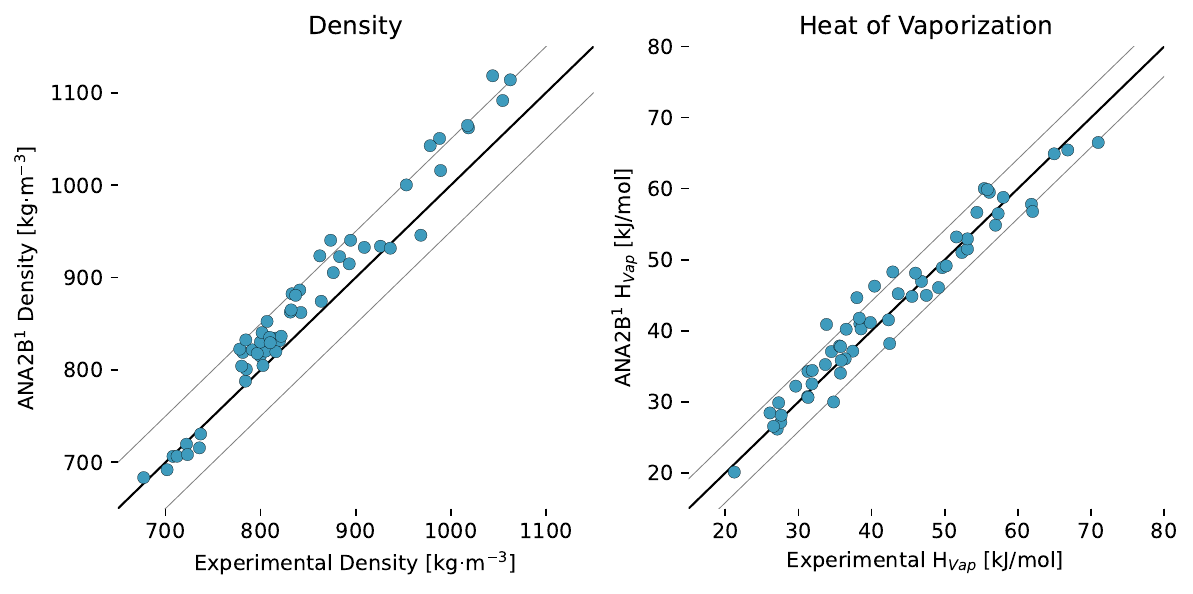}
    \caption{Results for condensed-phase properties of $57$ molecules used in the parametrization and validation of GROMOS 2016H66: density (left) and heat of vaporization (right).
    Black lines indicate equality $\pm$~50~kg/m$^{3}$ and $\pm$ ~4.184~kJ/mol.}
    \label{fig:results_md}
\end{figure}

Here, we rely on a dataset that was used to parametrize and validate the GROMOS 2016H66 FF~\cite{2016H66}. This dataset consists of a diverse set of $57$ small molecules and several properties including the heat of vaporization, density ($\rho$), isothermal compressability ($\kappa$), thermal expansion coefficient ($\alpha$), and the dielectric permittivity ($\epsilon$). We limit the analysis to the heat of vaporization and the density in this study due to the slow convergence of the other properties. Results with ANA2B$^1$ are shown in Table~\ref{tab:gromos_properties}. For both properties, we observe RMSE values comparable to the fixed-charge FF (IPA and GROMOS 2016H66) shown in Figure~\ref{fig:results_md} and Table~\ref{tab:gromos_properties}, confirming the observation made for the prediction of lattice energies, i.e., transferability to the condensed phase is possible for the ANA2B$^1$ model. These results are particularly noteworthy as GROMOS 2016H66 was parametrized on these two properties. The slightly smaller error of IPA for the density might stem from the fact that its parametrization included molecular crystals, indicating that the prediction of densities could be improved by incorporating condensed-phase structures during training. 
Finally, we note that as the only exception, two of three simulations of ethylenediamine in the liquid phase crashed after $24.3$ and $31.6$\,ps, respectively, with the ANA2B$^1$ model.

\begin{table}[H]
    \caption{Root-mean-square error (RMSE) for pure liquid properties of 57 systems used in the calibration and validation of the GROMOS 2016H66 FF \cite{2016H66}. Values for GROMOS 2016H66 and IPA were taken from the referenced publications. The uncertainty is given as the mean standard deviation obtained over four replica.
    More detailed error statistics is provided in Table S11 in the Supporting Information. The individual numerical values are given in Table S12.
    } 
    \centering
    \small
    \begin{tabular}{l | c c c}
    \hline
    Property & IPA\cite{MLFF}  & GROMOS 2016H66 \cite{2016H66} & ANA2B$^1$\\\hline
    H$_\text{vap}$\, [kJ/mol]& 4.5 & 3.5 & 2.8 $\pm$ 0.9\\
    $\rho$\, [kg/m$^{3}$] & 26.3 & 32.4 & 33.9 $\pm$ 6.2\\\hline
    \end{tabular}
    \label{tab:gromos_properties}
\end{table}

\subsection{Crystal Structure Prediction}
Having established a level of accuracy in the previous sections, this last section is concerned with the application of the ANA2B$^{\infty}$ model to the (retrospective) ranking of molecular crystals. As targets we use the structures, which were part of the CSP blind tests 3~\cite{CSP3BT}, 5~\cite{CSP5BT}, and 6~\cite{CSP6} organized by the Cambridge Crystallographic Data Centre in the past. These blind tests were chosen due to the availability of all submitted candidates, allowing for the least biased assessment of the ability to find the experimental crystal structure given a list of candidates. We limit ourselves to the pure and neutral targets restricted to H, C, N, O, F, S, Cl. Target XX of the third blind test was excluded due to convergence issues. For the third and fifth blind test, a ranking based on lattice energies is used. 
For the sixth blind test, we furthermore explore how additional contributions, such as entropic terms, impact the ranking.

\subsubsection{CSP Blind Tests 3 and 5}
\begin{figure}[H]
    \includegraphics[width=0.99\textwidth]{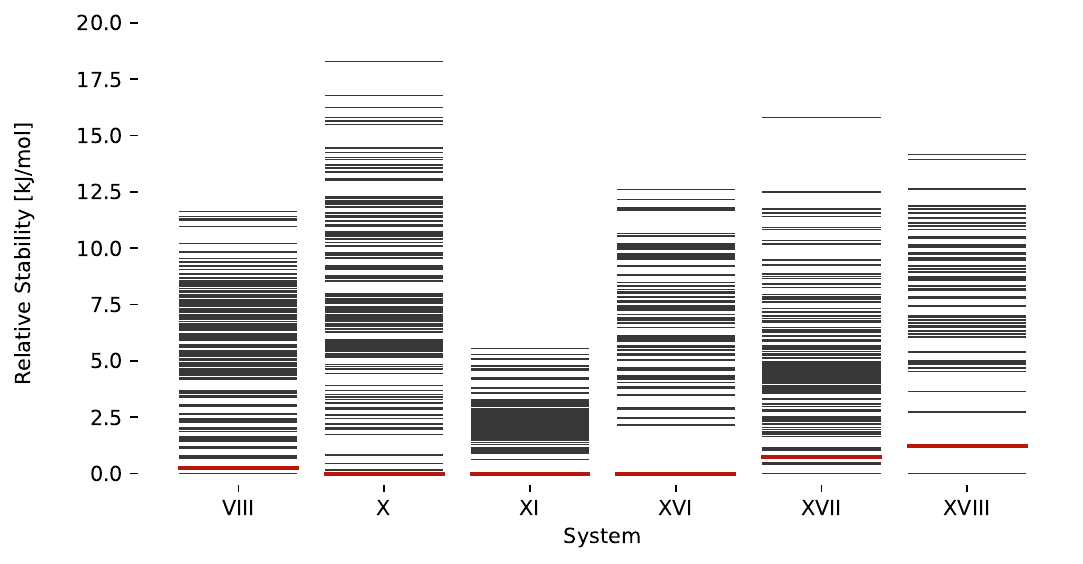}
    \caption{Stability ranking for the crystal structure for the compounds of the CSP blind tests 3 (VIII, X, XI)~\cite{CSP3BT} and 5 (XVI, XVII, XVIII)~\cite{CSP5BT} using the lattice energy predicted with the ANA2B$^{\infty}$ model. Each horizontal bar represents the stability of a structure with respect to the most stable structure. Red bars indicate experimental structures. The candidate structures were taken from the corresponding publications~\cite{CSP3BT, CSP5BT}.}
\label{fig:CSPS15}
\end{figure}

Rankings for targets stemming from the third and fifth blind test are shown in Figure~\ref{fig:CSPS15}. Candidates for blind test three submitted by van Eijck were generated using random search~\cite{vanEijck1}. Candidates for the fifth blind test submitted by Neumann \textit{et al.} were generated using Monte Carlo parallel tempering~\cite{TFF1}.
In all cases, a match with the experimental structure (red) would have been found as the most stable structure of within a window of $<1.3$\,kJ/mol.
Overall, these results underscore the strength of the proposed ML-augmented FF, which yields rankings that are in most cases comparable to rankings based on much more expensive methods such as system-tailored FFs~\cite{TFF2} or DFT~\cite{CSPXDM}.

\subsubsection{CSP Blind Test 6}
\begin{figure}[H]
    \includegraphics[width=0.9\textwidth]{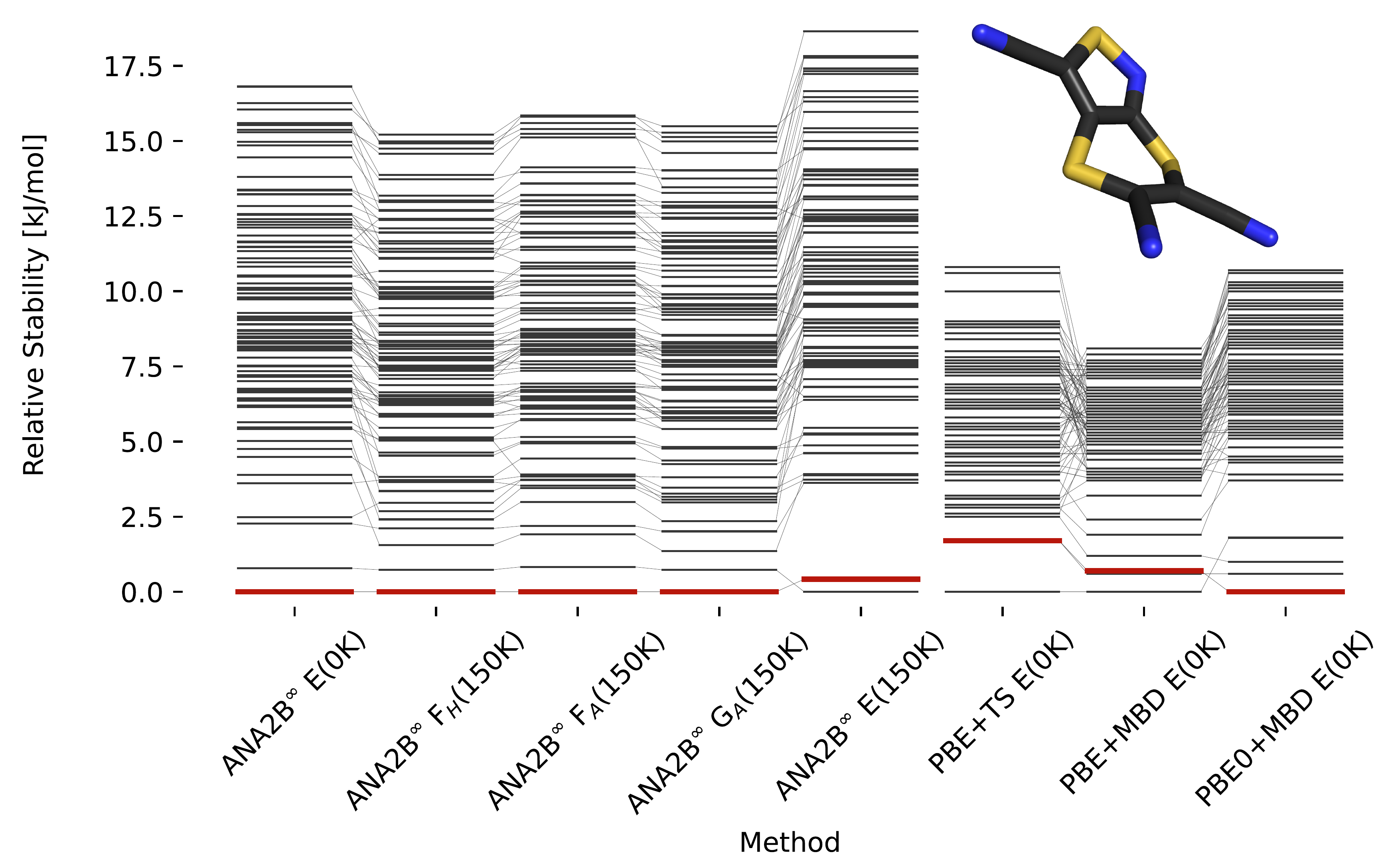}
    \caption{Stability ranking for the crystal structure of compound XXII. Each horizontal bar represents the stability of a structure with respect to the most stable structure. The stability is given in kJ/mol per molecule. Candidate structures and rankings for dispersion corrected PBE and PBE0 are taken from Ref.~\cite{HojaCSP}. Experimental structures are marked in red.}
\label{fig:XXII}
\end{figure}

\begin{figure}[H]
    \includegraphics[width=0.9\textwidth]{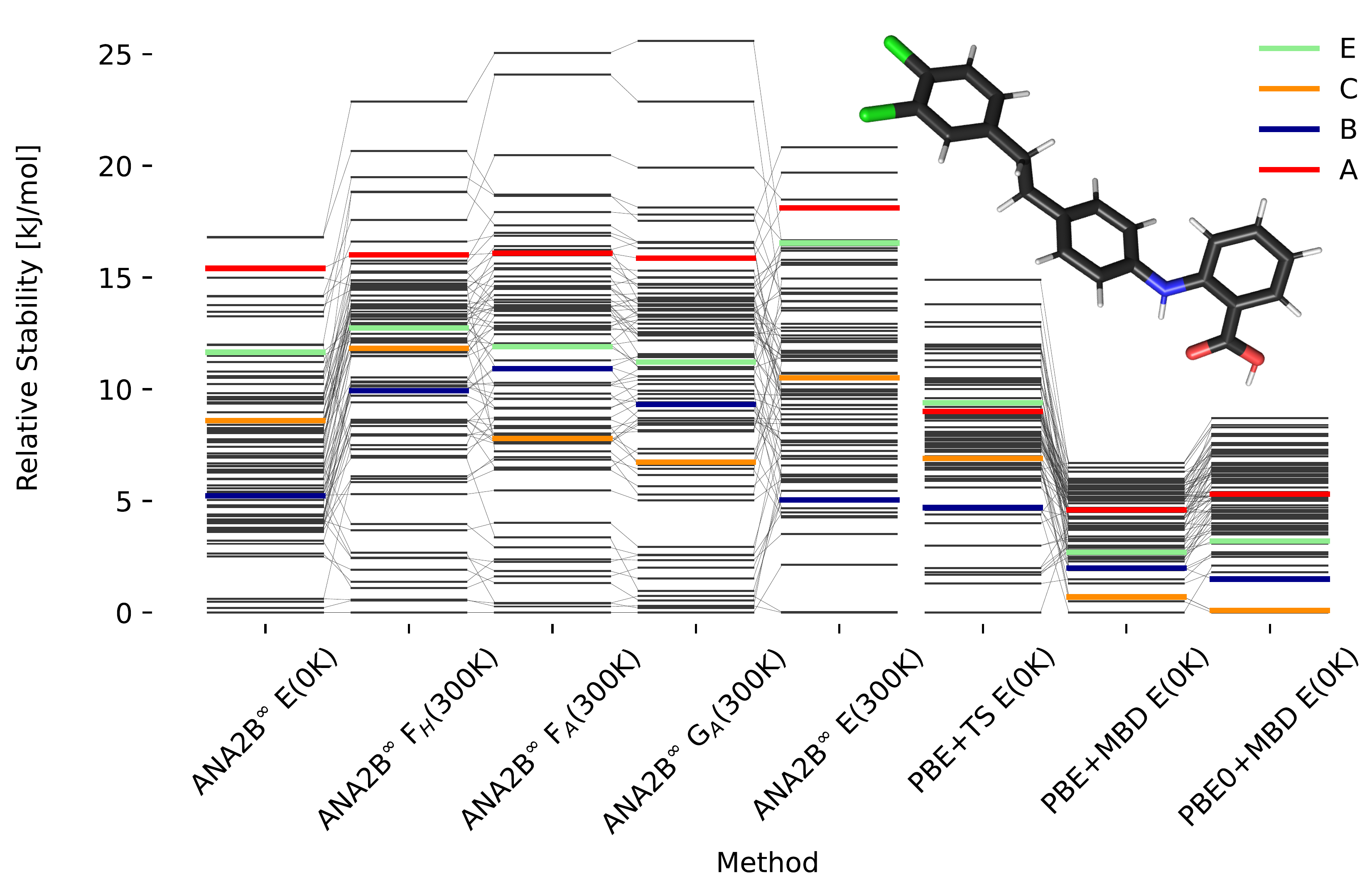}
    \caption{Stability ranking for the crystal structure of compound XXIII. Each horizontal bar represents the stability of a structure with respect to the most stable structure. The stability is given in kJ/mol per molecule. Candidate structures and rankings for dispersion corrected PBE and PBE0 are taken from Ref.~\cite{HojaCSP}. Experimental structures are marked in color.}
\label{fig:XXIII}
\end{figure}

\begin{figure}[H]
    \includegraphics[width=0.9\textwidth]{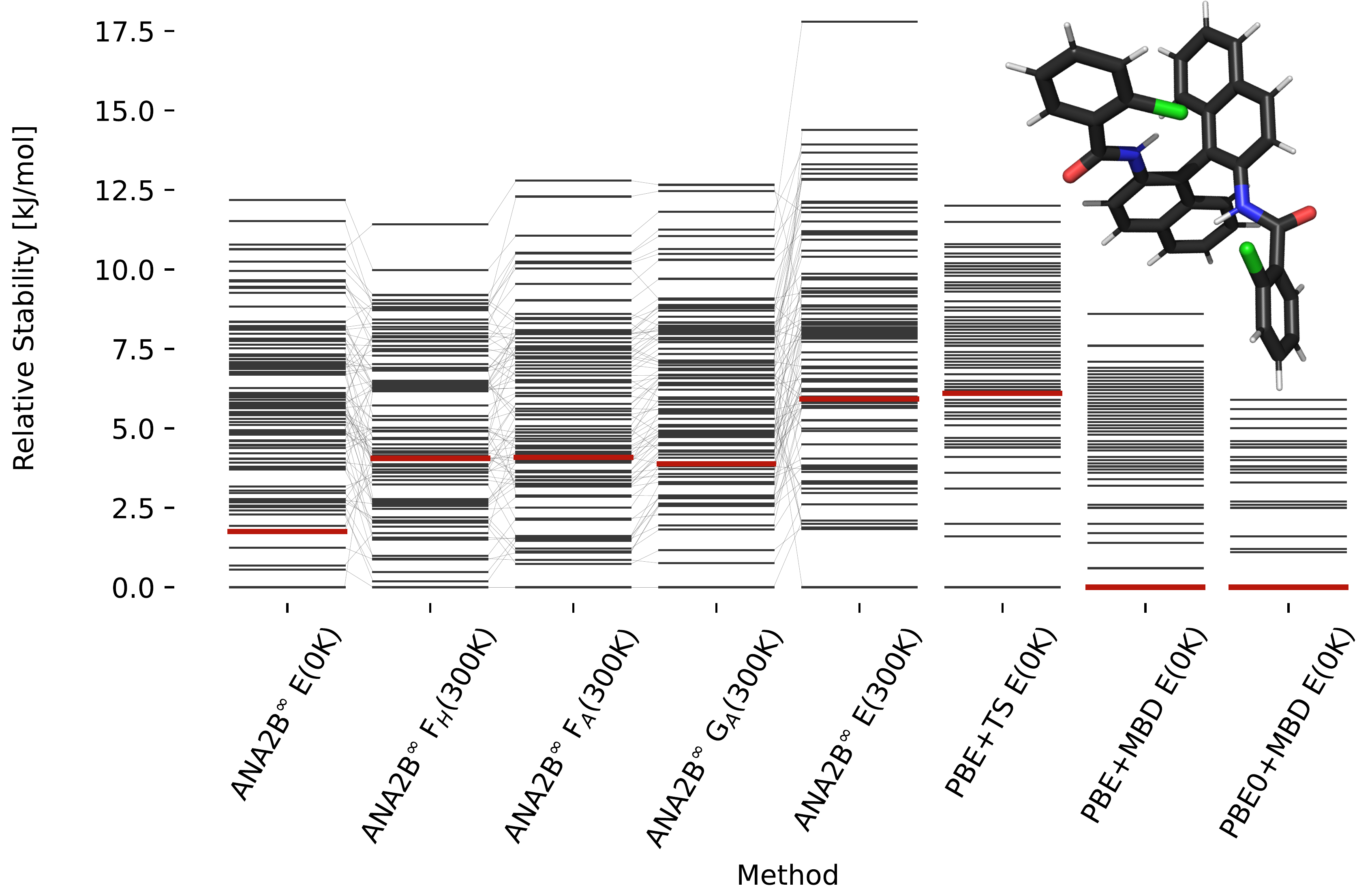}
    \caption{Stability ranking for the crystal structure of compound XXVI. Each horizontal bar represents the stability of a structure with respect to the most stable structure. The stability is given in kJ/mol per molecule. Candidate structures and rankings for dispersion corrected PBE and PBE0 are taken from Ref.~\cite{HojaCSP}. Experimental structures are marked in red. Note that PBE0+MBD is not available for all polymorphs of this structure.}
\label{fig:XXVI}
\end{figure}

In previous work, Hoja \textit{et al.}~\cite{HojaCSP} presented a workflow to rank crystal structures of the $6$th CSP blind test~\cite{CSP6} in a hierarchical manner. They generated candidate structures first using the tailor-made FF developed by Neumann and co-workers~\cite{TFF1, TFF2}, and subsequently ranked them with increasingly accurate methods, including vibrational contributions in the final ranking. Here, we base our study on the candidate structures made available as part of their work~\cite{HojaCSP}, which includes all known experimental structures. The exhaustive computational study by Hoja \textit{et al.} has provided insight into the different contributions stemming from DFT on different levels of theory and vibrational contributions, which we can use for a comparison with our ANA2B$^{\infty}$ model. Rankings for the three pure systems XXII, XXIII, and XXVI are shown in Figures \ref{fig:XXII}-\ref{fig:XXVI} based on the lattice energy (ANA2B$^\infty$ E(0K)), the harmonic Helmholtz free energy (ANA2B$^\infty$ F$_\text H$(T)), the Helmholtz free energy including anharmonic corrections (ANA2B$^\infty$ F$_\text A$(T)), the Gibbs free energy (ANA2B$^\infty$ G$_\text A$(T), and the mean potential energy during a molecular simulation (ANA2B$^\infty$ E(T)). Rankings for dispersion corrected PBE and PBE0 are taken from Ref.~\cite{HojaCSP}.

\begin{table}[H]
    \caption{Relative deviations in percentage ($(\text{pred.} - \text{exp.}) / \text{exp.}) \cdot 100\%$) from the experimental lattice cell parameters and volumes for the polymorphs minimized with ANA2B$^\infty$ and mean absolute percentage errors (MAPE).}
    \centering
    \small
    \begin{tabular}{l | c c c c c c c}
    \hline
        Systems & a & b & c &  $\alpha$ & $\beta$ & $\gamma$ & Volume\\\hline
        XXII-N2 & -0.78 & -0.49 & 1.11 & - & 0.76 & - & -0.66\\
        XXIII-A & -1.85 & -2.45 & 0.37 & - & -1.48 & - & -3.69\\
        XXIII-B & 2.55 & -0.49 & -4.92 & 3.97 & 1.56 & -1.49 & -3.47\\
        XXIII-C & -2.70 & -0.81 & -0.92 & 2.03 & 2.06 & 0.23 & -3.96\\
        XXIII-D & -2.20 & 0.63 & 0.99 & - & 2.07 & - & -2.47\\
        XXVI-N1 & -1.72 & -1.26 & -2.64 & 3.30 & 0.73 & 0.86 & -4.41\\\hline
        MAPE & 1.97 & 1.02 & 1.83 & 3.10 & 1.44 & 0.86 & 3.11\\\hline

    \end{tabular}
    
    \label{tab:crystal_params}
\end{table}

For compounds XXII and XXVI, the ANA2B$^\infty$ lattice energy ranks the experimental polymorph as the most stable (XXII) and the fifth most stable (XXVI) structure within a window of $2$\,kJ/mol. 
Interestingly, we do not observe a distinct benefit for the inclusion of corrections to the lattice energy based on entropic contributions. While in some cases a destabilization of non-experimental structures is observed, no clear improvement of the actual ranking is found.
This surprising finding suggests that improving the accuracy of the predicted energy might be the highest priority for future work. A fine-tuning on high-quality data of crystalline energies and/or gradients could be a possible solution. Such a fine-tuning might be particularly important for systems where a fine balance between intramolecular and intermolecular interactions exists, i.e., most flexible molecules.

A second interesting observation concerns compound XXIII, where the ANA2B$^\infty$ model fails to rank the experimental structures near the most stable candidate. This failure is most evident for polymorph A, which is in all cases ranked as one of the least stable structures. As the only exception, polymorph B is found within a window of a bit more than $5$\,kJ/mol. Importantly, several structures, most notably polymorph D and N70, could not be converged during the optimization or resulted in unstable MD simulations. In previous work~\cite{HojaCSP}, N70 was ranked as the most stable polymorph with PBE0+MBD+F$_\text{vib}$. 

Relative errors in percent of the lattice cell parameters with respect to the experimental structures are given in Table~\ref{tab:crystal_params}. A consistent underestimation of cell parameters and volumes is found, consistent with the results obtained for the densities of liquids. However, unlike for liquids sampled at finite temperatures, the underestimation of cell volumes might be explained partially with the optimization of cell parameters at $0$\,K.


\section{Conclusion}
In the present work, we have introduced a hybrid classical/ML potential for the simulation of molecular systems. Our work demonstrates that the combination of classical potentials with specific ML-based corrections can result in highly accurate, interpretable, and transferable potentials. 
The classical description of atomic interactions can thereby profit from augmentation with ML while ML can profit from the constraints imposed by classical models, especially for long-range electrostatics. The proposed hybrid approach could thus fill the existing methodological gap with a method, which can reach the accuracy of DFT at a computational cost between classical FF and semi-empirical methods while simultaneously improving the applicability of ML potentials.
In the present work, particular attention was given to the development of a ML-based approach, which can be used for condensed-phase systems but does not require reference data of such systems. However, the results for the crystal structure prediction indicate that the inclusion of some high-quality reference data of condensed-phase systems might be needed to fine tune the balance between intramolecular and intermolecular interactions. 

Besides improving the efficiency and computational cost, possible avenues for future investigations could include the explicit treatment of three-body interactions with a ML potential or higher-order polarization. However, both of these options would result in significant additional computational costs. 
An alternative route might be the application within a semi-empirical model instead of a classical FF.
In principle, the proposed pairwise ML potential could be applied to semi-empirical methods. Assuming that semi-empirical methods are able to accurately describe long-range interactions, a short-range pairwise potential might be able to largely resolve the limitations of semi-empirical models. This application might be particularly interesting for systems for which the classical approximations assumed in this work are not valid. In a similar vein, the pairwise potential could also be used to improve the description interactions between the QM and MM particles in QM/MM simulations, which typically still rely on classical Lennard-Jones potentials.

Overall, we anticipate that the proposed methods will significantly facilitate the parametrization of highly accurate FF. 

\section*{Software and Data Availability}
All datasets used in this work are publicly available (see the corresponding references in the text). The dataset used to train the intramolecular potential is published as part of this work and can be found on the ETH Research Collection \url{https://www.research-collection.ethz.ch/handle/20.500.11850/626683}
Code and model weights necessary to reproduce the results in this work are made available on GitHub: \url{https://github.com/rinikerlab/ANA2B}.

\section*{Acknowledgements}
The authors thank Felix Pultar for helpful discussions.

\printbibliography

\end{document}